\documentclass[12pt]{article}
\usepackage{graphicx}
\usepackage{epsf}
\usepackage{amssymb}
\usepackage{epsfig}
\def\dis{\displaystyle}
\def\beq{\begin{equation}}
\def\eeq{\end{equation}}
\def\be{\begin{equation}}
\def\ee{\end{equation}}
\def\ba{\begin{array}}
\def\ea{\end{array}}
\def\bt{\begin{tabular}}
\def\et{\end{tabular}}
\def\bea{\begin{eqnarray}}
\def\eea{\end{eqnarray}}
\def\beas{\begin{eqnarray*}}
\def\eeas{\end{eqnarray*}}
\def\bef{\begin{figure}}
\def\eef{\end{figure}}

\def\noi{\noindent}

\newcommand{\nc}{\newcommand}
\nc{\dy}{\displaystyle}

\nc{\pr}{{\rm I}} %
\nc{\se}{{\rm II}} %
\nc{\w}{{\rm w}} %
\nc{\s}{{\rm s}}
\nc{\um}{{1\over2}} %
\nc{\Hc}{{{\cal H}_{\rm c}}}

\begin{document}


\title{\vspace{-4truecm} {}\hfill{\small } \\
\vspace{1truecm}
Generalized Ginzburg-Landau models for non-conventional superconductors}%
\author{S. Esposito${}^{a,}$\footnote{sesposito@na.infn.it} \ and G. Salesi${}^{b,}
$\footnote{salesi@unibg.it}
\\
\footnotesize ${}^{a}$ Dipartimento di Scienze Fisiche, Universit\`a di Napoli
``Federico II'' \\ \footnotesize \& I.N.F.N. Sezione di Napoli, \\
\footnotesize Complesso Universitario
di M. S. Angelo, Via Cinthia, 80126 Naples, Italy, \\
\footnotesize ${}^{b}$ Facolt\`a di Ingegneria, Universit\`a Statale di Bergamo,
\\ \footnotesize
viale Marconi 5, 24044 Dalmine (BG), Italy \\ \footnotesize \& I.N.F.N. Sezione di
Milano, \footnotesize via G. Celoria 16, I-20133 Milan, Italy }

\maketitle

\begin{abstract}
\noi We review some recent extensions of the Ginzburg-Landau model able to describe
several properties of non-conventional superconductors. In the first extension, $s$-wave
superconductors endowed with two different critical temperatures are considered, their
main thermodynamical and magnetic properties being calculated and discussed. Instead in
the second extension we describe spin-triplet superconductivity (with a single critical
temperature), studying in detail the main predicted physical properties. A thorough
discussion of the peculiar predictions of our models and their physical consequences is
as well performed.
\end{abstract}


\section{Introduction}

\noi The Higgs mechanism \cite{Higgs,Bailin}, plays a basic role in the macroscopic
theory of gapped BCS superconductors, or in the corresponding Ginzburg-Landau (GL)
effective theory, where it accounts for the emergence of short-range electromagnetic
interactions mediated by massive-like photons (responsible, for example, of the Meissner
effect) \cite{Tinkham}. In the GL theory this is achieved by means of a complex order
parameter $\phi$, which can be interpreted as the wave function of the Cooper pair in its
center-of-mass frame. The classical phenomenological GL approach entails a (unique)
critical temperature $T_C$, without assuming a particular temperature-dependence of the
coefficient $a(T)$ appearing in the effective free energy function for unit volume,
expanded up to the $|\phi|^4$ order: \be F \simeq F_{\rm n} + a(T)|\phi|^2 +
\frac{\lambda}{4}\,|\phi|^4\,. \label{Effe} \ee Quantity $F_{\rm n}$ indicates the
normal-phase (not superconducting) free energy density; \  while $\lambda$, giving the
strength of the Cooper pair binding, is assumed to be approximately constant. Ginzburg
and Landau only assumed that the coefficient $a(T)$ is positive above $T_C$, vanishes
when the temperature approaches the critical value and becomes negative for $T<T_C$;
around the critical temperature changes very smoothly: \ $a(T) \simeq
\dot{a}(T_C)(T-T_C)$\,. In the alternative quantum field approach analyzed below we
instead adopt well-defined analytic expressions for $a(T)$ as a function of the
temperature.

Actually, because of the interaction of the charged scalar field $\phi$ with the
electromagnetic field $A^\mu$, the order parameter is usually associated to the Higgs
field responsible of the U(1) spontaneous symmetry breaking (SSB)
\cite{Higgs,Bailin,Tinkham} occurring during the normal state-superconducting phase
transition. As a consequence of the symmetry breaking, due to a non-vanishing expectation
value of the order parameter in the ground state below the critical temperature, the
photon acquires a mass (causing the Meissner effect) and the system becomes
superconducting. By adopting this approach, we can initially start from a
relativistically invariant Lagrangian containing the interaction of a single $\phi$ with
$A^\mu$ as well as the $\lambda$ self-interaction (hereafter $\hbar=c=1$):
\begin{equation}
{\cal L}=\left( D_{\mu }\phi \right)^{\dagger }\left( D^{\mu }\phi \right) +
m^2\phi^\dagger\phi - \frac{\lambda}{4}(\phi^\dagger\phi)^2
-\frac{1}{4}F_{\mu\nu}F^{\mu\nu}\,, \label{Elle}
\end{equation}
where \ $m^2>0$, \ $F_{\mu\nu}\equiv\partial_\mu A_\nu - \partial_\nu A_\mu$ is the
electromagnetic field strength, \ and \ $D_{\mu }\equiv\partial_{\mu }+2ieA_{\mu }$ \ is
the covariant derivative ($2e$ is the electric charge of a Cooper pair). Given the above
Lagrangian, the effective free energy density at finite temperature is formally identical
to the GL expression given in (\ref{Effe}). However, despite the outstanding importance
of the GL theory in superconductivity, as well as in other physical systems, it has still
not been solved exactly beyond the mean-field approximation. Whilst this was not a
serious problem for traditional superconductors, where the Ginzburg temperature interval
is small around the critical temperature, the situation has changed especially with the
advent of high-$T_c$ superconductors. In fact, for these systems, the Ginzburg
temperature interval is large and we may expect strong field fluctuations and critical
properties beyond the mean-field approximation. Indeed, in high-$T_c$ superconductors
several experiments have observed critical effects in the specific heat \cite{critical},
although the presence of a magnetic field generally makes the situation more complicated.
On a theoretical side, the effect of gauge field fluctuations causes great difficulties
in the critical phenomena theory and, unlike the simpler $\phi^4$ theory for a neutral
superfluid, the exact critical behaviour remains unknown. It is well-known that at the
mean-field level the superconductive transition is discontinuous, but it seems that this
result is confirmed even when field fluctuations are included \cite{fluct}. This is also
confirmed by numerical simulations of lattice models for small values of the
Ginzburg-Landau parameter $\varkappa$, while for large $\varkappa$ the results are
consistent with a continuous, second-order phase transition \cite{second}. It is thus a
general belief that the standard GL model leads to a first-order transition instead of a
continuous transition, but several other studies at one-loop (and even at two-loop)
approximation have been carried out in recent years (see, for example, \cite{one1, one2,
one3, one4} and references therein). Some of them entail runaway solutions of the GL
equations (pointing towards first-order transitions), while others find a scaling
behaviour with a new stable fixed point in the space of static parameters. Also, again
beyond the plain mean-field approximation, Kleinert has shown the existence of a
tricritical point in a superconductor, by taking the vortex fluctuations into account.

In the present paper we review some extensions of the GL model, recently proposed
\cite{TwoTC, Term2TC, Magn2TC, monophase} in order to describe the properties of
non-conventional superconductors. In a first model, discussed in the next section, we
describe $s$-wave superconductors endowed with two (slightly) different critical
temperatures, and study in detail both the thermodynamical and the magnetic properties of
such systems. Instead in section 3 we report on a straightforward generalization of the
standard GL model accounting for spin-triplet superconductivity (with a single critical
temperature), again studying the main physical features of the medium considered.
Finally, in section 4 we summarize and discuss the relevant results obtained.

\section{A GL-like model for $s$-wave superconductors with two critical temperatures}

\subsection{The model}

\noi Let us start by expanding a complex field $\phi$ as follows
\begin{equation}
\phi \equiv \frac{1}{\sqrt 2}(\eta_0+\eta)\,{\rm e}^{i\theta/\eta_0}\,, \label{Phi_I}
\end{equation}
where $\eta_0$ is a real constant, $\eta$ and $\theta$ are real fields. Then, if we let
the scalar field fluctuate around the minimum of the free energy, a condensation of the
field $\eta$ takes place as a result of the $U(1)$ SSB. In Eq.\,(\ref{Phi_I}) the
constant field $\eta_0/\sqrt{2}$ is defined as the expectation value (the condensation
value) of the modulus of the scalar field $\phi$. \ Finite-temperature one-loop quantum
corrections to the $T=0$ expression of the free energy density lead to \cite{NXC} \be
F_\pr = F_{\rm n} + \um a_\pr(T)\eta_0^2 + \frac{\lambda}{16}\,\eta_0^4 \ee with \be
a_\pr = - m^2 + \frac{\lambda + 4e^2}{16}\,T^2\,. \label{a_I} \ee The parameter $a_\pr$
vanishes when the temperature approaches a critical value given by \be T_1 =
2\sqrt{\frac{4m^2}{\lambda + 4e^2}}\,. \label{T1} \ee Below $T_1$ the expectation value
of $\eta_0^2$ which minimizes the free energy function results to be \be \eta_0^2(T) =
-\frac{4a_\pr(T)}{\lambda}\,. \label{etaquad} \ee Alternatively, we may expand the field
$\phi$ as:
\begin{equation}
\phi \equiv \frac{1}{\sqrt 2}(\phi_0 + \phi_{a} +i \phi_{b}), \label{Phi_II}
\end{equation}
where $\phi_0$ is a real constant, and $\phi_{a}, \phi_{b}$ are two real scalar fields.
Now we assume that a condensation takes place in the field $\phi_a$ (or, equivalently, in
$\phi_b$) rather than in the component $\eta$. In Eq.\,(\ref{Phi_II}) the constant field
$\phi_0/\sqrt{2}$ is defined as the  expectation value of the real part of $\phi$. In
this case, after such condensation, the effective Helmholtz energy density writes \be
F_\se = F_{\rm n} + \um a_\se(T)\phi_0^2 + \frac{\lambda}{16}\,\phi_0^4 \ee with
\cite{Bailin} \be a_\se = - m^2 + \frac{\lambda + 3e^2}{12}\,T^2\,. \label{a_II} \ee From
the vanishing of $a_\se$ we now derive a \emph{different} critical temperature \be T_2 =
2\sqrt{\frac{3m^2}{\lambda + 3e^2}}\,. \label{T2} \ee Since \ $\infty>\lambda>0$, \ we
correspondingly have \ $\displaystyle \frac{\sqrt{3}}{2}\,T_1<T_2<T_1$. \ Accordingly,
for very large self-interaction, \ $\lambda/e^2\to\infty$, \ \emph{we predict a maximum
difference of} 15\% \emph{between the two critical temperatures} \cite{TwoTC}.

Below $T_2$ the expectation value for $\phi_0^2$ which minimizes the free energy function
is given by \be \phi_0^2(T) = -\frac{4a_\se(T)}{\lambda}\,. \label{phiquad} \ee We
understand the appearing of a new lower critical temperature when expanding the
exponential in Eq.\,(\ref{Phi_I}) in $\theta/\eta_0$ and comparing with
Eq.\,(\ref{Phi_II}):
\begin{eqnarray}
\phi_0 & \sim & \eta_0
\nonumber \\
\phi_a & \sim & \eta - \frac{\theta}{2}\left(\frac{\theta}{\eta_0}\right)+ \cdots\label{A5}\\
\phi_b & \sim & \theta \, + \, \eta\left(\frac{\theta}{\eta_0}\right)+ \cdots\,.
\nonumber
\end{eqnarray}
The degrees of freedom carried out by the real scalar fields $\phi_a ,\phi_b$ are
different from those corresponding to $\eta, \theta$, and tend to coincide only in the
limit $\eta_0\rightarrow \infty$. Actually, in Eqs.\,(\ref{A5}) the higher orders in
$\eta_0^{-1}$ contribute at the denominator of the expression (\ref{T1}) as an additional
$\lambda/3$ term; that is an increased effective self-interaction of the Cooper pairs
arises ($\lambda\rightarrow \lambda_{\rm eff}=4\lambda/3$) \cite{TwoTC}.

Since, as we have seen, two different condensations are allowed to occur inside the same
system, we do not a priori exclude any of them. Hence we are led to introduce two order
parameters, that is two scalar charged fields: the first one related to the condensation
of the modulus of $\phi_\pr$ (the corresponding phase will be hereafter denominated as
``phase I''); while the second one related to the condensation of the real part of
$\phi_\se$ (``phase II'').

Neglecting possible interactions between the two scalar fields, the total Lagrangian now
writes:
    \be
    {\cal L} =\left(D_{\mu }\phi_\pr \right)^{\dagger }\left( D^{\mu }\phi_\pr
    \right) + m^2\phi_\pr^\dagger\phi_\pr - \frac{\lambda}{4}(\phi_\pr^\dagger\phi_\pr)^2+
    \left(D_{\mu }\phi_\se \right)^{\dagger }\left( D^{\mu }\phi_\se
    \right) + m^2\phi_\se^\dagger\phi_\se - \frac{\lambda}{4}(\phi_\se^\dagger\phi_\se)^2 -
    \frac{1}{4}F_{\mu\nu}F^{\mu\nu}\,.
    \label{Ellex2}
    \ee
As a matter of fact, starting from high values and then lowering the temperature we meet
a first SSB at the critical temperature $T_1$: the medium becomes superconducting. Since
the II-phase term $a_\se(T){\phi_0}^2 + \lambda\,{\phi_0}^4$ in the free energy density
is negative for $T<T_2$, by further lowering the temperature at $T=T_2$ the condensation
involving the second order-parameter is energetically favored and a new (second-order)
phase transition starts. Below $T_2$ the system is ``more'' superconducting with respect
to the GL standard case since, in addition to the phase-I Cooper pairs, we should observe
also the formation of phase-II Cooper pairs. Such two superconducting phases correspond
to different condensations of electrons in Cooper pairs which exhibit different
self-interaction, and are described by different scalar fields. The realization of one of
the two regimes is ruled by the relative strength of the Cooper pair self-interaction
($\lambda$) with respect to the electromagnetic interaction ($e$).

Correspondingly, the total free energy density, being an additive quantity, results as
the sum of contributions from normal-conducting electrons, phase-I superconducting Cooper
pairs, and phase-II superconducting Cooper pairs:
\begin{eqnarray}
F = F_{\rm n} & \ \mbox{\rm for} \ & T>T_1\,,
\label{Norm}\\
F = F_{\rm n} + \um a_\pr(T){\eta_0}^2 + \frac{\lambda}{16}\,{\eta_0}^4 & \ \mbox{\rm
for} \ & T_2<T<T_1\,, \ \ \
\label{I}\\
F = F_{\rm n} + \um a_\pr(T){\eta_0}^2 + \frac{\lambda}{16}\,{\eta_0}^4 +
\um a_\se(T){\phi_0}^2 + \frac{\lambda}{16}\,{\phi_0}^4 & \ \mbox{\rm for} \ & T<T_2\,,
\label{I+II}
\end{eqnarray}
($\eta_0$ indicates the expectation value of $|\phi_\pr|$; $\phi_0$ indicates the
expectation value of ${\rm Re}\{\phi_\se\}$).

From Eqs.(\ref{T1}) and (\ref{T2}) we are able to put the two free parameters of our
theory, i.e. the ``mass squared'' $m^2$ and the self-interaction coupling constant
$\lambda$ as functions of the two critical temperatures: \be m^2 =
\frac{e^2T_1^2T_2^2}{4(4T_2^2-3T_1^2)}\,, \ee \be \lambda =
\frac{12e^2(T_1^2-T_2^2)}{4T_2^2-3T_1^2}\,. \ee Therefore experimental measurements of
$T_1$ and $T_2$ could yield an estimate of the dynamical parameters ruling the SSB and
the electron binding in Cooper pairs. Notice that such a goal is not possible in the
framework of the standard GL, theory where the parameters in (\ref{Effe}) are not
explicitly determined.

\subsection{Thermodynamical properties}

\noi Inserting the expressions obtained above in (\ref{a_I}), (\ref{etaquad}),
(\ref{a_II}), and (\ref{phiquad}), also the expectation values of the two scalar fields
can be expressed in terms of $T_1$ and $T_2$: \be \eta_0^2(T) =
\frac{T_2^2(T_1^2-T^2)}{12(T_1^2-T_2^2)}\,, \ee \be \phi_0^2(T) =
\frac{T_1^2(T_2^2-T^2)}{12(T_1^2-T_2^2)}\,. \ee By inserting in (\ref{I}) and
(\ref{I+II}) we may compare, for $T<T_2$, the behavior of the free energy in the GL case,
where (\ref{I}) holds also for $T<T_2$, and in the case of two-phases superconductors for
which, instead, (\ref{I+II}) applies. The free energy difference results to be \be \Delta
F \equiv F_{\rm GL} - F_{\rm 2ph} =
\frac{e^2T_1^4(T_2^2-T^2)^2}{48(4T_2^2-3T_1^2)(T_1^2-T_2^2)}\,. \label{Delta F} \ee We
see that such a difference increases by lowering the temperature and reaches its maximum
for $T=0$.

The pressure is given by \ $\displaystyle P=-\left.\frac{\partial{\cal F}}{\partial
V}\right|_T$, \ where ${\cal F}=FV$ is the free energy. Since that the superconductive
part of the free energy density is independent of the volume, we have \be \Delta P \equiv
P_{\rm GL} - P_{\rm 2ph} = - \Delta F < 0\,. \ee Hence the pressure is expected to be
larger for two-phases superconductors. Thus the differences in the free energy and in the
pressure become more sensible far from $T_2$, near to absolute zero.

From \ $\displaystyle S = -\left.\frac{\partial F}{\partial T}\right|_{V}$, \ for
$T<T_2$, we get the difference in the entropy density: \be \Delta S \equiv S_{\rm GL} -
S_{\rm 2ph} = \frac{e^2T_1^4T(T_2^2-T^2)}{12(4T_2^2-3T_1^2)(T_1^2-T_2^2)}\,. \ee Being
$\Delta S>0$, we can say that the two-phases superconductors are in a sense more
``ordered'' than the GL ones, the maximum difference for the entropy being reached at \
$T=T_2/\sqrt{3}$\,.

\begin{figure}
\centering
\includegraphics[scale=0.7]{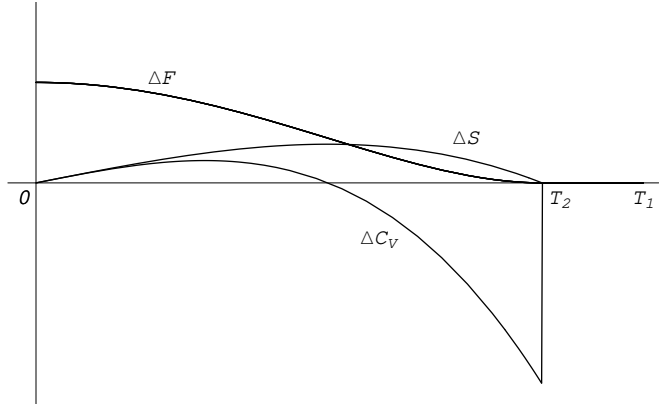}
\caption{Differences between GL and two-phases superconductors} 
\end{figure}

We can compare as well the latent heat absorbed during the formation of the
superconducting phase in GL and two-phase superconductors at a given temperature $T<T_2$
($S_0$ indicates the entropy of the normal-phase) \be \lambda_{\rm GL}(T) = T(S_0 -
S_{\rm GL}) = \frac{e^2T_2^4T^2(T_1^2-T^2)}{12(4T_2^2-3T_1^2)(T_1^2-T_2^2)}\,\,; \ee \be
\lambda_{\rm 2ph}(T) = T(S_0 - S_{\rm 2ph}) = \frac{e^2T^2[T_2^4(T_1^2-T^2) +
T_1^4(T_2^2-T^2)]}{12(4T_2^2-3T_1^2)(T_1^2-T_2^2)}\,. \ee The difference between
$\lambda_{\rm GL}$ and $\lambda_{\rm 2ph}$ reaches its maximum at $T=T_2/2$.

Finally, applying the well-known formula for the specific heat at constant volume \be C_V
= T\,\left.\frac{\partial S}{\partial T}\right|_V\,, \ee we obtain the difference in
$C_V$ ($T<T_2$) \be \Delta C_V \equiv C_{V_{\rm GL}} - C_{V_{\rm 2ph}}
=\frac{e^2T_1^4T(T_2^2-3T^2)}{12(4T_2^2-3T_1^2)(T_1^2-T_2^2)}\,, \ee which is positive
for \ $0<T<T_2/\sqrt{3}$, \ negative for \ $T_2/\sqrt{3}<T<T_2$, \ and vanishes at \
$T=T_2/\sqrt{3}$. \ As it is seen in the figure, whilst $\Delta F(T_2)$ and $\Delta
S(T_2)$ vanish, so that $F$ and $S$ are continuous in $T_2$, quantity $\Delta C_V(T_2)$
is not zero: \be \Delta C_V(T_2) =
-\,\frac{e^2T_1^4T_2^3}{6(4T_2^2-3T_1^2)(T_1^2-T_2^2)}<0\,. \ee We then observe a finite
jump in the specific heat also in the transition from the  superconducting first phase
(I) to the second one (II), while in GL superconducting media only one discontinuity is
expected (for $T=T_1$, when the system changes from the normal to the superconducting
regime). This sudden change in the heat capacity is a distinguishing characteristic of a
first order phase transition. Since the jump of the specific heat in $T=T_1$ results to
be \be \Delta C_V(T_1) =-\,\frac{e^2T_2^4T_1^3}{6(4T_2^2-3T_1^2)(T_1^2-T_2^2)}\,, \ee the
ratio between the two discontinuities can be written as \be \frac{\Delta C_V(T_2)}{\Delta
C_V(T_1) } = \frac{T_1}{T_2}\,. \ee Being the above ratio larger than 1 (and smaller than
$\sqrt{4/3}$\,\,), we expect the two jumps to be comparable. Thus also the second jump at
the lower temperature ---just a novel effect because it happens between two
superconducting phases--- could be experimentally investigated and measured. Notice also
that, as expected, both discontinuities increase indefinitely in the large Cooper pairs
self-interaction limit, \ $\lambda/e^2\to \infty$, \ $(T_1/T_2)^2\to 4/3$.

\subsection{Magnetic properties}

\noi Let us start with the {\em Meissner effect}, namely the rapid decaying to zero of
magnetic fields in the bulk of a superconductor. The distance from the surface beyond
which the magnetic field vanishes is known as the \textit{London penetration depth}, and
can be written in terms of the effective (after SSB
and Higgs mechanism \cite{Naddeo, Tinkham, Higgs}) photon mass%
$$ %
\delta = \frac{1}{m_A} = \sqrt{\frac{1}{8e^2|\phi_{\rm min}(T)|^2}} \,,
$$ %
where $\phi_{\rm min}(T)$ is the expectation value of the field in the minimum energy
state at a given temperature.  Exploiting Lagrangian (\ref{Ellex2}), in the phase-I
($T_2<T<T_1$) we
therefore have %
\be \delta_\pr = \sqrt{\frac{1}{8e^2|\eta_0(T)|^2}}\,,
\ee %
whilst in the phase-II ($T<T_2$) we now have two contributions to
the photon mass %
\be \delta_\se = \sqrt{\frac{1}{8e^2(|\eta_0(T)|^2 + |\chi_0(T)|^2)}}\,.
\ee %
The expectation values $\eta_0(T)$ and $\chi_0(T)$ can be expressed \cite{Term2TC} as
functions of the critical temperatures \be \eta_0^2(T) = -\frac{2\, a_\w(T)}{\lambda} =
\frac{T_2^2(T_1^2-T^2)}{24(T_1^2-T_2^2)}\,, \ee %
\be \chi_0^2(T) = -\frac{2 \, a_\s(T)}{\lambda} = \frac{
T_1^2(T_2^2-T^2)}{24(T_1^2-T_2^2)}\,. \ee As a consequence we can write the London
penetration lengths as follows:
\begin{equation}
\delta_\pr =  \left[\frac{e^2 T_2^2T_1^2}{3
(T_1^2-T_2^2)}\left(1-\frac{T^2}{T_1^2}\right)\right]^{-\um}
\end{equation}
for the phase-I; and
\begin{equation}
\delta_\se = \left\{\frac{e^2 T_2^2T_1^2}{3
(T_1^2-T_2^2)}\left[\left(1-\frac{T^2}{T_1^2}\right)
+\left(1-\frac{T^2}{T_2^2}\right)\right]\right\}^{-\um}
\end{equation}
for the phase-II. Let us stress that, below the second critical temperature $T_2$, the
penetration length of the magnetic field is smaller with respect to the GL one-phase
superconductors.

The \textit{coherence length}
$$ %
\xi=  \frac{1}{m_\phi(T)}
$$ %
measures the distance over which the scalar field varies sensitively and is related to
the mean binding length of the electrons in a Cooper pair. The coherence length can be
expressed (via Higgs mechanism \cite{Naddeo, Higgs, Tinkham}) as a function of the
temperature-dependent effective mass of the scalar field which results from a quantum
field calculation including one-loop radiative correction: namely
\begin{equation}
m_{\phi_\w}(T) = \sqrt{- {a_\w(T)}}
\end{equation}
for the weak field ($T<T_1$), and \be m_{\phi_\s}(T) = \sqrt{- {a_\s(T)}} \ee for the
strong field ($T<T_2$). As expected, for $T\neq 0$,
$$ %
m_{\phi_\w}^2 = m^2 - \frac{\lambda + 4e^2}{16}\, T^2 \,>\, m_{\phi_\s}^2 = m^2 -
\frac{\lambda + 3e^2}{12}\,T^2
$$ %
since the (negative) binding energy between the electrons is larger for the
strongly-coupled Cooper pairs.

By applying the above definition we get \textsl{two different coherence lengths} for the
two fields
\begin{equation}
\xi_\w(T) = \frac{1}{m_{\phi_\w}(T) }\,,
\end{equation}
\begin{equation}
\xi_s(T) = \frac{1}{m_{\phi_\s}(T)}\,.
\end{equation}
At absolute zero the renormalized masses are equal to the bare mass $m$ and the two
coherence lengths reduce to the common value:
\begin{equation}
 \xi_0 = \frac{1}{m} = 2\frac{\sqrt{ (4 T_2^2-3T_1^2)}}{e T_1T_2}\,.
 \label{p0}
\end{equation}
Let us write explicitly the temperature dependence of the coherence lengths for the two
types of Cooper pairs :
\begin{eqnarray}
\xi_\w(T) &=&  \frac{1}{m_{\phi_\w}(T)}=\frac{\xi_0}{\dy\sqrt{1
-\left(\frac{T}{T_1}\right)^2}},
\qquad\quad \mbox{for } T<T_1\,, \label{p1} \\
\xi_\s(T) &=&  \frac{1}{m_{\phi_\s}(T) }=\frac{\xi_0}{\dy\sqrt{1 -\left(\frac{
T}{T_2}\right)^2}}, \qquad\quad  \mbox{for } T<T_2\,. \label{p2}
\end{eqnarray}

\noindent By increasing the intensity of the magnetic field entering type-I
superconductors, when $H$ reaches a critical value $\Hc$, perfect diamagnetism and
superconductivity are suddenly destroyed through a first-order phase transition. The
critical magnetic field measures the ``condensation energy", given by the difference
between the free energies of the normal and superconducting states
$$ %
F - F_{\rm n} = - \um\mu_0\Hc^2 \,.
$$ %
Exploiting the above equation we obtain the critical field in the phase-I, for
$T_2<T<T_1$:
\begin{equation}
\Hc^\pr = \sqrt{\frac{2}{\mu_0\lambda}}\,|a_\w| = \frac{e T_2^2(T_1^2-T^2)}{2\sqrt{6\mu_0
(4T_2^2-3T_1^2)(T_1^2-T_2^2)}}\,,
\end{equation}
while in the phase-II, for $T< T_2$, we have
\begin{equation}
\Hc^\se = \sqrt{\frac{2}{\mu_0\lambda}\left(a_\w^2+a_\s^2\right)} =
\Hc^\pr\sqrt{1+\frac{T_1^4(T_2^2-T^2)^2}{T_2^4(T_1^2-T^2)^2}}\,.
\end{equation}
Let us notice that there is no discontinuity at $T_2$: \ $\Hc^\pr(T_2)= \Hc^\se(T_2)$;
but, while for $T\lesssim T_1$ the critical field  decreases linearly, for $T\lesssim
T_2$ we have instead a quadratic behavior
\begin{equation}
\Hc^\se \simeq \Hc^\pr(T_2)\left[ 1 +
\frac{2T_1^4}{T_2^2(T_1^2-T^2)^2}(T_2-T)^2\right]\,.
\end{equation}

\noindent For type-II superconductors there exist two different critical fields, $\Hc_1$,
the \textsl{lower critical field}, and $\Hc_2$ the \textsl{upper critical field}. We
observe perfect diamagnetism only applying a field $H<\Hc_1$ whilst, when
\mbox{$\Hc_1<H<\Hc_2$}, non superconducting vortices can arise in the bulk of the medium.
Abrikosov \cite{L9, Abrikosov} showed that a vortex consists of regions of circulating
supercurrent around a small central normal-metal core: the magnetic field is able to
penetrate through the sample inside the vortex cores, and the circulating currents serve
to screen out the magnetic field from the rest of the superconductor outside the vortex.

In the present model there are actually two different coherence lengths for two different
types of Cooper pairs. Correspondingly, in the phase-II we shall have \textsl{different
upper and lower critical fields} in ``domains" of the sample occupied by Cooper pairs of
the same type, either weakly-coupled or strongly-coupled. As a consequence, in a given
domain we can have (or not) Abrikosov vortices depending on the type of scalar field
condensed in that domain. Such an inhomogeneity of the spatial distribution of the
vortices in two-phase superconductors should result in a net, detectable difference with
respect to GL superconductors: in principle, for $T<T_2$, a section of the material
should show vortical and non-vortical sectors, by contrast to the homogeneous
distribution of vortex cores in ordinary superconductors.

Now, by starting from the London equation we can easily obtain the explicit expression of
the lower critical field \cite{Annett, L9}
$$ %
\Hc_1 = \frac{\Phi_0}{4\pi\mu_0\delta^2}\ln{\left(\frac{\delta}{\xi}\right)}
$$ %
where $\Phi_0\equiv \dy\frac{\pi}{e}$ is the so-called quantum magnetic flux unit. Thus,
in the phase-I, the lower critical field writes:
\begin{equation}
\Hc_1^\pr =
\frac{\Phi_0}{4\pi\mu_0\delta_\pr^2}\ln{\left(\frac{\delta_\pr}{\xi_\w}\right)} =
h_1\left(1-\frac{T^2}{T_1^2}\right)
\end{equation}
where %
\be h_1\equiv \frac{\Phi_0}{4\pi\mu_0} \, \frac{e^2T_1^2T_2^2}{3(T_1^2-T_2^2)} \,
\ln\left[\sqrt{\frac{6(T_1^2-T_2^2)}{4T_2^2-3T_1^2}}\right] ,.
\ee %
As said above, in the phase-II we have two distinct $\Hc_1^\se$:
\be %
\Hc_{1_\w}^\se =
\frac{\Phi_0}{4\pi\mu_0\delta_\se^2}\ln{\left(\frac{\delta_\se}{\xi_\w}\right)} =
h_1\left[\left(1-\frac{T^2}{T_1^2}\right)+ \left(1-\frac{T^2}{T_2^2}\right)\right]
\left\{1-\frac{\ln\left[1 +
\frac{T_1^2(T_2^2-T^2)}{T_2^2(T_1^2-T^2)}\right]}{\ln\left[\frac{6(T_1^2-T_2^2)}{4T_2^2-3T_1^2}\right]}\right\}
\ee %
in the weak-field domains; and
\be %
\Hc_{1_\s}^\se =
\frac{\Phi_0}{4\pi\mu_0\delta_\se^2}\ln{\left(\frac{\delta_\se}{\xi_\s}\right)} =
h_1\left[\left(1-\frac{T^2}{T_1^2}\right)+ \left(1-\frac{T^2}{T_2^2}\right)\right]
\dy\left\{1-\frac{\ln\left[1 +
\frac{T_2^2(T_1^2-T^2)}{T_1^2(T_2^2-T^2)}\right]}{\ln\left[\frac{6(T_1^2-T_2^2)}{4T_2^2-3T_1^2}\right]}\right\}
\ee in the strong-field domains.

Let us now pass to the upper critical field which can be expressed as follows
\cite{Annett, L9}
$$ %
\Hc_2 = \frac{\Phi_0}{2\pi \mu_0}\frac{1}{\xi^2}\,.
$$ %
In the weak-field ($T<T_1$) and strong-field ($T<T_2$) domains the above equation writes,
respectively:
\begin{eqnarray}
\Hc_{2_{\rm w}} &=&
\frac{\Phi_0}{2\pi\mu_0}\frac{1}{\xi_\w^2}=h_2\left(1-\frac{T^2}{T_1^2}\right),
\qquad\quad  \mbox{for } T<T_1\,, \\
\Hc_{2_{\rm s}} &=&
\frac{\Phi_0}{2\pi\mu_0}\frac{1}{\xi_\s^2}=h_2\left(1-\frac{T^2}{T_2^2}\right),
\qquad\quad  \mbox{for } T<T_2\,,
\end{eqnarray} %

\be h_2 \equiv \frac{\Phi_0}{\pi\mu_0} \, \frac{e^2
T_1^2T_2^2}{(4T_2^2-3T_1^2)}\,. \ee %

\

\noindent For $T\rightarrow 0$ we have:
\begin{equation}
\Hc_{2_{\rm w}} = \Hc_{2_{\rm s}} = h_2\,,
\end{equation}
while, for $T \to T_2$:
\begin{equation}
\Hc_{2_{\rm w}} \neq \Hc_{2_{\rm s}} = 0.
\end{equation}

\begin{figure}
\centerline{\hbox{\psfig{figure=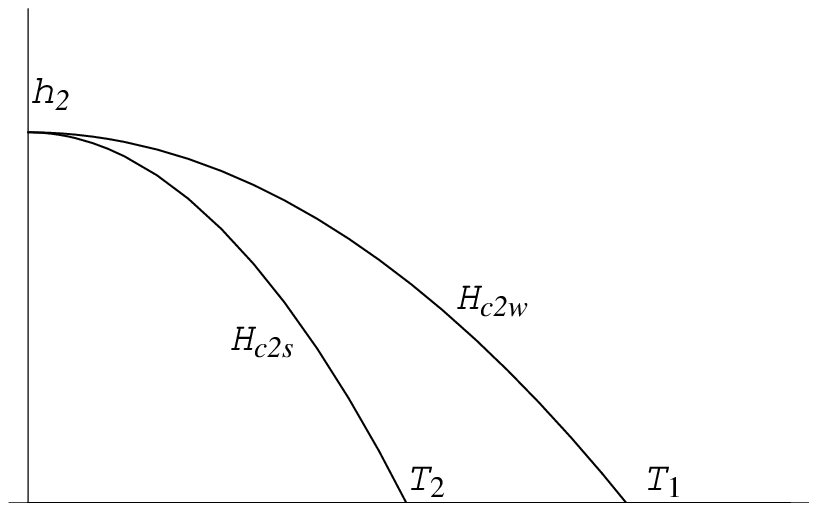,width=0.45\textwidth}}} \caption{Upper critical
fields vs. temperature for weak-field and strong-field domains} \label{fig1}
\end{figure}

The well-known dimensionless \textsl{Ginzburg-Landau parameter}
$$ %
\kappa \equiv \frac{\delta}{\xi}
$$ %
determines whether a medium is a type-I or type-II superconductor: for $\kappa<1/\sqrt 2$
it is a type-I superconductor; whilst for $\kappa>1/\sqrt 2$ it is a type-II
superconductor.

In the phase-I we find:
\begin{equation}
\kappa_{\,\pr} = \sqrt{\frac{6(T_1^2-T_2^2)}{4T_2^2-3T_1^2}}
\end{equation}
which is independent of temperature in agree with the GL theory.  On imposing the
condition $\kappa_{\,\pr}<1/\sqrt 2$ we infer that the material, for $T_2<T<T_1$, is a
type-I superconductor if $\dy1\leq\left(\frac{T_1}{T_2}\right)^2\leq \frac{16}{15}$; and
a type-II superconductor if $\dy\frac{16}{15}\leq\left(\frac{T_1}{T_2}\right)^2\leq
\frac{4}{3} $.

In the phase-II we easily get: \be \kappa_{\se{\rm w}} =
\left[1+\frac{T_1^2(T_2^2-T^2)}{T_2^2(T_1^2-T^2)}\right]^{-\um}\!\!\!\!\kappa_{\,\pr} \ee
for the weak-field domains; and \be \kappa_{\se{\rm s}} =
\left[1+\frac{T_2^2(T_1^2-T^2)}{T_1^2(T_2^2-T^2)}\right]^{-\um}\!\!\!\!\kappa_{\,\pr} \ee
for the strong-field domains.

The dependence of $\kappa_\se$ on temperature could be observed as a net deviation from
the  temperature-independent behavior of GL superconductors. Notice that
$\kappa_\se<\kappa_\pr$: therefore a type-II superconductor for temperatures in the
region $T_2<T<T_1$ could become a type-I superconductor for $T<T_2$, if $\kappa$
decreases below  $1/\sqrt{2}$. Quantities $\kappa_{\se{\rm w}}$ and $\kappa_{\se{\rm s}}$
turn out to be equal for $T\sim 0$: \be \kappa_{\se{\rm w}} = \kappa_{\se{\rm s}} \simeq
\frac{\kappa_{\,\pr}}{\sqrt 2}\,. \ee

\section{Generalized GL model for $p$-wave superconductors}

\noi If we introduce two {\em mutually interacting} order parameters, we will not in
general describe two-phases superconductors but, actually, \textit{spinning} Cooper pairs
and rotational degrees of freedom in superconductivity. In the Bardeen, Cooper, and
Schrieffer (BCS) theory for conventional superconductors, the electrons are paired into a
zero total angular momentum state, with zero spin and zero orbital angular momentum: $J =
L = S = 0$. As a matter of fact, in BCS superconductors the $s$-wave is shown to
correspond to the minimum energy state with maximum attraction between the electrons in a
Cooper pair. Indeed, soon after the BCS theory was advanced, Kohn and Luttinger
\cite{Kohn} predicted that, if the mutual interaction is \textit{repulsive} in all
partial wave channels, the Cooper pairs result to be bonded by a weak residual attraction
(out of the Coulomb repulsion) in higher angular momentum channels: this is the so-called
\textit{Kohn-Luttinger effect}. On the other hand it exists also a $p$-wave
Cooper-pairing in superfluid $^3$He (which is, as said above, the liquid counterpart of
GL superconductors). Actually, we can meet $p$-wave superconductivity in certain
``heavy-electron" compounds (\textit{heavy fermion systems} as, e.g., UPt$_3$) and in
special materials recently discovered as, e.g., Sr$_2$RuO$_4$ \cite{Maeno} which is the
only known metal oxide displaying $p$-wave superconductivity. Let us recall that the
$p$-wave Cooper pairs are always spin-triplets ($S$=1) because of Pauli's exclusion
principle applied to systems composed of a pair of particles endowed with odd ($L$=1)
total orbital quantum number. Taking into account this property, we shall now put forward
a simple GL-like model just for spin-triplet superconductors.

\subsection{The model}

\noi Let us consider the physical system described by one doublet of complex scalar
fields $\phi_1, \phi_2$ through the following Lagrangian density:
\begin{eqnarray}
{\cal L}&=&-\frac{1}{4}F_{\mu \nu }F^{\mu \nu }+\left( D_{\mu
}\phi_1 \right)^{\star }\left( D^{\mu }\phi_1 \right)
+ \left( D_{\mu }\phi_2 \right)^{\star }\left( D^{\mu }\phi_2 \right)-\lambda (|\phi_1
|^{2}-\frac{1}{2}\phi_{0}^{2})^2 \nonumber \\ & & - \lambda(|\phi_2|^{2}
-\frac{1}{2}\phi_{0}^{2})^2 + V(\phi_1,\phi_2). \label{2a}
\end{eqnarray}
Here the covariant derivative $D_\mu\equiv\partial_\mu+ieA_\mu$ describes the minimal
electromagnetic interaction of the two scalar fields, while the first term in the
Lagrangian (with $F_{\mu \nu }\equiv \partial_\mu A_\nu -\partial_\nu A_\mu$) accounts
for the kinetic energy of the free electromagnetic field $A_\mu$. The complete potential
term for the interaction of the two scalar fields is composed of three different terms, \
$V=V_{\phi A}+ V_{\rm self}+ V(\phi_1, \phi_2)$, \ the first two of them describing the
usual interaction between the electromagnetic field and the charged scalar field (coming
from the covariant derivative), and $V_{\rm self}$ ruling the self interaction of the
scalar fields: $V_{\rm self}\equiv\lambda |\phi_1 |^{4}+ \mu |\phi_2|^{4}$. For the
interaction between the two scalar fields we instead adopt the following nonlinear term:
\begin{equation} \label{v12}
V(\phi_1,\phi_2)\equiv
-\frac{\lambda\phi_0^4}{8}\ln^2{\frac{\phi_1^{\vphantom{\star}}}{\phi_1^\star}
\frac{\phi_2^\star}{\vphantom{\phi_2^\star}\phi_2^{\vphantom{\star}}}}.
\end{equation}
Let us study the small fluctuations of the two scalar fields around the  minimum of the
energy corresponding to $\phi_1=\phi_2=\phi_0/\sqrt{2}$ by expanding both scalar fields
as follows:
\begin{eqnarray}
&&\phi_1 \equiv \frac{1}{\sqrt
2}(\phi_0+\eta_1)\,{\rm e}^{i\theta_1/\phi_0}, \label{phi1}\\
&&\phi_2 \equiv \frac{1}{\sqrt 2}(\phi_0+\eta_2)\,{\rm e}^{i\theta_2/\phi_0},
\label{phi2}
\end{eqnarray}
where $\eta_1$, $\eta_2$, $\theta_1$, $\theta_2$ are real fields. From these definitions,
the above interaction term can be written more simply as follows \be V(\phi_1,\phi_2) =
\frac{\lambda\phi_0^2}{2}(\theta_1-\theta_2)^2\,. \ee Notice that $V(\phi_1,\phi_2)$ is
positive-definite, then describing a \textit{repulsion} between the two fields with
strength $\lambda\phi_0^2$ equal to the mass squared $m_W^2$ (see below). Note also that
$V(\phi_1,\phi_2)$ corresponds to the main term of the expansion for small phase
differences \cite{Legget} of the Legget interaction \ $\gamma(\phi_1^\star\phi_2 +
\phi_1\phi_2^\star)$.

By inserting Eqs.\,(\ref{phi1},\ref{phi2}) into the Lagrangian density (\ref{2a}) and
performing the gauge transformation: $A_{\mu }\rightarrow A_{\mu }+\partial_{\mu }\Lambda
$ with
\begin{equation}
\Lambda \equiv -\frac{1}{2e\phi_0}(\theta_1+\theta_2)\,, \label{5a}
\end{equation}
we obtain the following Lagrangian, up to quadratic terms in the fields:
\begin{eqnarray}
{\cal L}&\simeq&-\frac{1}{4}F_{\mu \nu }F^{\mu \nu }+e^{2}\phi_{0}^2A_\mu A^\mu
+\frac{1}{2}\partial_{\mu }\eta_1\partial^{\mu }\eta_1+\frac{1}{2}\partial_{\mu
}\eta_2\partial^{\mu }\eta_2
+ \frac{1}{2}\partial_{\mu}(\theta_1 - \theta_2)\partial^{\mu}(\theta_1 - \theta_2)
\nonumber\\ & & +\lambda\phi_{0}^{2}\eta_1^2+ \lambda \phi_{0}^{2}\eta_2^2+\frac{\lambda
\phi_0^2}{2}(\theta_1 - \theta_2)^2. \label{12a}
\end{eqnarray}
Let us set
\begin{equation}
\eta_3\equiv\frac{1}{\sqrt{2}}(\theta_1 - \theta_2)\label{13a}
\end{equation}
and define the triplet field $W_a\equiv(\eta_1,\eta_2,\eta_3)$. The Lagrangian describing
our physical system now becomes:
\begin{eqnarray}
{\cal L}&\simeq&-\frac{1}{4}F_{\mu \nu }F^{\mu \nu }+ m_A^{2}A_\mu A^\mu
+\frac{1}{2}(\partial_{\mu }W_a)(\partial^{\mu }W_a)+m_W^{2}W_aW_a\label{14a}
\end{eqnarray}
with
\begin{equation}
m_A^2= e^{2}\phi_{0}^2 , \qquad \quad m_W^2=\lambda\phi_{0}^{2}.\label{15a}
\end{equation}
As a result, only one of the original four degrees of freedom embedded into two charged
(complex) scalar fields is disappeared, by giving rise to a massive photon as in the
standard GL model. By virtue of the interaction potential in Eq.\,(\ref{v12}), the
remaining three degrees of freedom all have the same mass, and can thus be combined to
form a triplet field $W$ (i.e. a triplet spinor representation of SU(2)), suitable to
describe a so-called $p$-wave superconductor. We stress that, notwithstanding the
simultaneous condensation of two real degrees of freedom, the key point in our model is
the particular interaction term we have introduced, which prevents a gauge transformation
to re-absorb one more degree of freedom (only the \textit{sum} of the phases of the
complex fields turns out to be ``eaten", but not even the \textit{difference}). Such a
very peculiar interaction breaks the isotropy of the original medium and allows pairs of
electrons to arrange into possible $S=1$ (instead of $S=0$) Cooper pairs. As a matter of
fact, the emergence of a triplet field is a signal of the occurred ``anisotropization''
of the system, which can no more be described by a singlet scalar field.

\subsection{Predicted properties for triplet superconductors}

\noi The order parameter describing $p$-wave superconductors may be associated in our
model to the above triplet Higgs field $W_a$ which is responsible of the U(1) spontaneous
symmetry breaking occurring during the normal state-superconducting-phase transition.
Therefore, from the Lagrangian (\ref{2a}), the effective free energy density at finite
temperature $T$, resulting from the quantum fields calculation, including one-loop
radiative corrections \cite{Bailin,NXC}, is given by
\begin{eqnarray}
F(T) &=& F_{\rm n}(T) + a(T)|\phi_1|^2 +a(T)|\phi_2|^2
+\lambda|\phi_1|^4 + \lambda|\phi_2|^4 +
a(T)\frac{\phi_0^2}{8}\left|\ln{\frac{\phi_1^{\vphantom{\star}}}{\phi_1^\star}
\frac{\phi_2^\star}{\vphantom{\phi_2^\star}\phi_2^{\vphantom{\star}}}}\right|^2,
\end{eqnarray}
where \be a(T) = - m_W^2 + \frac{\lambda + e^2}{4}\,T^2\,, \label{za_I} \ee label $n$
referring to the normal (non superconducting) phase. The coefficient $a$ vanishes when
the temperature approaches a critical value given by \be T_c =
\sqrt{\frac{4m_W^2}{\lambda + e^2}}\,. \label{zT1} \ee Below $T_c$ the expectation values
of the scalar fields $\phi_1$ and $\phi_2$ which minimize the free energy function
results to be \be |\phi_1(T)| = |\phi_2(T)| = \sqrt{-\frac{a(T)}{2\lambda}}\, ,
\label{zetaquad} \ee while the third degree of freedom defined in Eq.\,(\ref{13a})
fluctuates around the zero  expectation value, corresponding to $\theta_1 = \theta_2$.
This last occurrence directly comes from the fact that the non-linear characteristic
potential term in Eq. (\ref{v12}) is non-negative definite, so that the minimum of the
free energy is reached when it vanishes. In this case, our model practically reduces to a
``simple'' doubling of the standard GL theory making recourse to two scalar order
parameters. As a consequence, it is very easy to re-obtain the usual main properties for
$p$-wave superconductors considered here.

The London penetration length of the magnetic field inside the superconductor arises due
to the presence of a massive photon, that is:
\begin{equation}
\delta = \frac{1}{m_A} = \frac{1}{e \phi_0} \, ,
\end{equation}
while the coherence length of the Cooper pairs described by the triplet scalar field is
given by:
\begin{equation}
\xi = \frac{1}{m_W} = \frac{1}{\phi_0 \sqrt{\lambda}} = \frac{\xi_0}{\dis\sqrt{1-
\frac{T^2}{T_c^2}}} \, .
\end{equation}
The \textit{critical magnetic field} $H_c$, measuring the condensation energy $F(T) -
F_{\rm n}(T) = - \mu_0 H_c^2 /2$ of the superconductor system can be obtained as follows:
\begin{equation}
H_c^2 = \frac{1}{\mu_0} \, \frac{a^2(T)}{\lambda} = H_{c0}^2 \left( 1- \frac{T^2}{T_c^2}
\right)^2 \,.
\end{equation}
By taking the derivative of the free energy function with respect to temperature, we
easily get the entropy gain with respect to the normal phase:
\begin{equation}
S-S_n = \frac{\partial \;}{\partial T} \left( - \frac{a^2(T)}{2\lambda} \right) = S_0
\left( 1- \frac{T^2}{T_c^2} \right) \frac{T}{T_c} \, .
\end{equation}
Finally, we can write down the expected discontinuity of the specific heat at the
critical point:
\begin{equation}
\Delta C_V = T \, \frac{\partial \;}{\partial T} \left( S - S_n \right) = S_0 \left( 1- 3
\, \frac{T^2}{T_c^2} \right) \frac{T}{T_c} .
\end{equation}

\section{Conclusion}

\noi In the framework of the GL theory, we have developed some models in order to
describe different physical systems experiencing two $S=0$ superconducting phases, whose
critical temperatures $T_1, T_2$ differ at most of $15\%$, or, alternatively, $p$-wave
superconductors by means of two mutually interacting order parameters which condensate
simultaneously at a same critical temperature.

In the first model, two different condensation regimes of two scalar fields with equal
(bare) mass and self-interaction strength arise, describing Cooper pairs formed by
differently interacting electrons, such different interaction arising from quantum loop
corrections. We have calculated the thermodynamical properties of such two-phase
superconductors, the most peculiar one being a second discontinuity in the specific heat,
and considered the main magnetic properties as well. Below the second critical
temperature, the penetration length of the magnetic field is smaller with respect to the
usual one-phase superconductors (even of about $70\%$), depending on temperature. This is
easily explained by the emergence of a second kind of Cooper pairs, in which electrons
are more bonded than in the first kind of pairs, leading also to two distinct behaviors
of the critical magnetic fields in the two superconducting phases. As a consequence, two
different coherence lengths for the electrons in the different Cooper pairs exist, this
resulting in peculiar superconductive properties. For instance, even if in the region
between $T_1$ and $T_2$ the system is a type-II superconductor, depending on the ratio of
the critical temperatures [$16/15\leq (T_1/T_2)^2\leq 4/3$], below $T_2$ it instead could
behave as a type-I superconductor if $\kappa$ decreases sufficiently ($\kappa$ becoming
$<1/\sqrt{2}$). Moreover, for $T<T_2$, the GL parameter $\kappa$ is not constant, and
exhibits a characteristic dependence on the temperature, a result strongly deviating from
the predictions of the GL theory. Perhaps all these effects have not yet been observed in
any material, due to the very small difference (no more than $15\%$) between the two
critical temperatures, but this seems to be not a really difficult task for dedicated
experiments, since those effects are very peculiar. The experimental investigation on the
novel kind of superconductors discussed here should then consist mainly in the careful
search for materials which may exhibit the particular properties of the two kinds of
electron pairs considered. The major merit of this model, with respect to the existing
theories for (usual and) unusual superconductors is its full predictability, since all
the basic physical properties are expressed in terms of the two measurable critical
temperatures, rather than in terms of unknown quantities such as self-interaction
coupling constants, quasi-particle effective masses, or mean distance between Cooper
electrons. Let us also remark that attractive interactions (``Cooper-effect") and
``gapped" energy spectrum characterize both quantum theory of {\em fermionic} superfluids
(as e.g. $^3$He) and BCS theory of superconductivity. Consequently we might expect that
the basic properties for two-phase superconductors could analogously apply to a sort of
``two-phase Fermi superfluids'' (endowed with two critical temperatures) as well.

In the second model, the standard GL theory has been generalized in order to describe
$p$-wave superconductors by means of two mutually interacting order parameters which
condensate simultaneously at the same critical temperature (since the $\lambda\phi^4$
self-interaction is the same for both fields). After the condensation we remain with
three massive degrees of freedom (in addition to a massive photon, related to the
Meissner-effect) which can be put in correspondence to the three components of a $S=1$
triplet mean-field describing spinning $p$-wave Cooper pairs. In this model the main
magnetic and thermodynamical properties (including the discontinuity in the specific
heat) of $p$-wave superconductors turn out to be essentially the same as for conventional
$s$-wave superconductors discussed above.

It is very intriguing that completely different systems may exhibit similar physical
properties, thus encouraging further theoretical and experimental studies in this
direction.

\end{document}